\documentclass[12pt,preprint]{aastex}

\begin{document}

%\slugcomment{Version 1.6}

\newcommand{\MtNC}{CH$_3$NC}
\newcommand{\MtCN}{CH$_3$CN}
\newcommand{\VyCN}{CH$_2$CHCN}
\newcommand{\VyNC}{CH$_2$CHNC}
\newcommand{\EtCN}{CH$_3$CH$_2$CN}
\newcommand{\EtNC}{CH$_3$CH$_2$NC}
\newcommand{\CYCN}{HC$_4$CN}
\newcommand{\CYNC}{HC$_4$NC}

\title{Interstellar Isomers: The Importance of Bonding Energy Differences}

\author{Anthony J. Remijan\altaffilmark{1,2}, J. M. Hollis\altaffilmark{1}, F. J. Lovas\altaffilmark{3},\\ 
D. F. Plusquellic\altaffilmark{3} \& P. R. Jewell\altaffilmark{4}}

\altaffiltext{1}{NASA Goddard Space Flight Center, Computational and Information Sciences and Technology Office, Code 606, Greenbelt, MD  20771}
\altaffiltext{2}{National Research Council Resident Research Associate}
\altaffiltext{3}{Optical Technology Division, National Institute of Standards 
and Technology, Gaithersburg, MD  20899}
\altaffiltext{4}{National Radio Astronomy Observatory, P. O. Box 2, 
Green Bank, WV  24944-0002}

\begin{abstract}

We present strong detections of methyl cyanide (CH$_3$CN), 
vinyl cyanide (CH$_2$CHCN), ethyl cyanide (CH$_3$CH$_2$CN) and cyanodiacetylene (HC$_4$CN) molecules
with the Green Bank Telescope (GBT) toward the Sgr B2(N) molecular cloud.  Attempts to detect the corresponding
isocyanide isomers were only successful in the case of methyl isocyanide (CH$_3$NC) for its  
$J_K=1_0-0_0$ transition, which is the first interstellar report of this line.
To determine the spatial distribution of CH$_3$NC, we used 
archival Berkeley-Illinois-Maryland Association (BIMA) array data for the $J_K=4_K-3_K$ ($K=0-3$) transitions but no emission 
was detected.  From ab initio calculations, the bonding 
energy difference between the cyanide and isocyanide molecules is $>$8500 cm$^{-1}$ ($>$12,000 K).  
Thus, cyanides are the more stable isomers and would likely be formed more preferentially over 
their isocyanide counterparts.  That we detect \MtNC\ emission with a single antenna 
(Gaussian beamsize($\Omega_B$)=1723 arcsec$^2$) but not with an interferometer ($\Omega_B$=192~arcsec$^2$), 
strongly suggests that CH$_3$NC has a widespread spatial distribution toward the Sgr B2(N) region. 
Other investigators have shown that \MtCN\ is present both in the LMH hot core of Sgr~B2(N) and in the surrounding 
medium, while we have shown that \MtNC\ appears to be deficient in the LMH hot core. Thus,
large-scale, non-thermal processes in the surrounding medium may account for the 
conversion of \MtCN\ to \MtNC\ while the LMH hot core, which is dominated by thermal processes, 
does not produce a significant amount of \MtNC. Ice analog experiments by other investigators have shown that 
radiation bombardment of \MtCN\ can produce \MtNC, thus supporting our observations.  
We conclude that isomers separated by such large bonding energy differences are distributed in
different interstellar environments, making the evaluation of column density ratios between 
such isomers irrelevant unless it can be independently shown that these species are co-spatial.

\end{abstract}


\hfill

\section{INTRODUCTION}
     
The observations of isomers in the interstellar medium is becoming increasingly more
common as new and larger molecules are being detected. Currently, 135 molecules
have been detected in interstellar and circumstellar environments.  Excluding
diatomic molecules, $\sim$30\% of all interstellar molecules have observed isomeric 
counterparts.  In addition, as the number of atoms in an interstellar molecule increases, so
do the number of isomers.  For example, of the 9 interstellar molecules containing 8 atoms, 
5 are isomers. With an empirical formula C$_2$H$_4$O$_2$,
acetic acid (CH$_3$COOH), glycolaldehyde (CH$_2$OHCHO) and methyl formate 
(CH$_3$OCHO) are isomers; and for H$_2$C$_6$, hexapentaenylidene
and triacetylene (HC$_6$H) are isomers.  Isomerism may provide
important clues to the formation of interstellar molecules.  As such, an important
class of isomers are cyanide (-CN) and isocyanide (-NC) species.

Hydrogen cyanide (HCN), methyl cyanide (CH$_3$CN), vinyl cyanide 
(CH$_2$CHCN), and ethyl cyanide (CH$_3$CH$_2$CN) all have large abundances and are
easily detectable in a number of interstellar environments and are readily seen
in the spectra of several molecular line surveys (e.g.\ Friedel et al.\
2004, Nummelin et al.\ 1998).  However, while there have been detections
of HNC in interstellar and cometary environments, there have been no confirmed
detection of the isocyanides \MtNC, \VyNC, and \EtNC.  The first tentative
detection of \MtNC\ was toward Sgr~B2(OH) by Cernicharo et al.\ (1988) from observations
of the $J = 4-3$, $J = 5-4$, and $J = 7-6$ transitions. Though the lines were
highly contaminated from other known interstellar species, they calculated an 
CH$_3$NC/CH$_3$CN abundance ratio of $\sim$0.03-0.05.  Since this initial claim, 
there have been no further observations to confirm the existence of interstellar \MtNC, 
in particular toward the LMH hot core of Sgr~B2(N), the best known source of large molecules.  Thus, in 
this work, we confirmed the existence of \MtNC\ and attempted to observe other large isocyanides 
by observing the same (or similar) low energy transitions of the cyanides \MtCN, \VyCN,
\EtCN, and \CYCN\ and their corresponding isocyanide isomers toward the 
the high-mass star forming region, Sgr~B2(N).

\section{OBSERVATIONS AND RESULTS}

The observations using the NRAO 100-m Robert C. Byrd Green Bank Telescope (GBT) 
were made in 2004 February 24, February 29 and March 29.  Table 1 lists the molecular line 
parameters of the transitions of the cyanides and isocyanides sought.
The species, calculated rest frequencies, transition quantum numbers, 
line strengths, upper level energies, and continuum levels are listed in the first six columns
of Table 1.  The GBT half-power beamwidth ($\theta_B$) is given as 
740$''$/$\nu$~[GHz].   The K-band receiver set used has a frequency range covering 
18 GHz~to~22.5 GHz. For the observations reported herein, the average beamwidth $<\theta_B>$=39$''$ 
corresponds to an average Gaussian beamsize of $<\Omega_B>$=$\pi \theta_B^2$/[4~ln(2)]~=~ 
1723~arcsec$^2$ and an average beam efficiency $<\eta_B>$$\sim$0.8.  The GBT spectrometer was 
configured in its 8 intermediate frequency (IF), 200 MHz, 3-level mode which permits
observing four 200 MHz frequency bands at a time in two polarizations through the use 
of offset oscillators in the IF.  This mode affords 24.4~kHz channel separation.  
Antenna temperatures are on the $T_A^*$ scale (Ulich \& Haas 1976) with estimated 
20\% uncertainties.  The Sgr~B2(N-LMH) J2000 pointing position employed was 
$\alpha$= 17$^h$47$^m$19$^s$.8, $\delta$ = -28$^o$22$'$17$''$ and the LMH
systemic LSR source velocity of +64~km~s$^{-1}$ was assumed.   Automatically updated dynamic 
pointing and focusing corrections were employed based on realtime temperature measurements 
of the structure input to a thermal model of the GBT; zero points were adjusted typically every 
two hours or less using the calibrators 1626-298 and/or 1733-130. Data were taken in the 
OFF-ON position-switching mode, with the OFF position 60$'$ east in azimuth 
with respect to the ON source position.  A single observation consisted of two 
minutes in the OFF source position followed by two minutes in the ON source 
position.  The two polarization outputs from the spectrometer were averaged in the 
final data reduction process to improve the signal-to-noise ratio.

Figure 1 shows our attempt to detect cyanide and isocyanide 
species toward Sgr~B2(N) with the GBT.  The panels in the left column show
the detections of \MtCN, \VyCN, \EtCN\ and \CYCN.  We have successfully detected 
all species in the left column and for the first time the 
$J_{K_a,K_c}=2_{11}-1_{10}$ transition of interstellar \EtCN\ is reported. 
In the panels containing the transitions of \MtCN,
\VyCN\ and \EtCN, the intensity scale (ordinate) decreases, showing that the emission
from the larger species is in general weaker.  The abscissa is the radial velocity
with respect to the local standard of rest (LSR) calculated for the rest frequency of the transition
shown at the top left of each panel assuming a source velocity of 
+64~km~s$^{-1}$.  Dashed lines show LSR velocities of +64, +73 and +82~km~s$^{-1}$.
Molecules with no hyperfine (HF) structure previously observed with the GBT such as 
propenal (CH$_2$CHCHO), propanal (CH$_3$CH$_2$CHO), and glycolaldehyde (CH$_2$OHCHO) 
show absorption components at +64 and +82~km~s$^{-1}$  and an emission component at +73~km~s$^{-1}$ 
(Hollis et al.\ 2004a, b).  In general, the cyanide spectra shown in the left panel of Figure 1 are consistent with
this source velocity structure.  In the cases of CH$_3$CN and CH$_2$CHCN,
however, the source velocity structure is confused by molecular HF
splitting.  The fiducial on each HF component denotes the position
corresponding to +64~km~s$^{-1}$.  Thus, it can be seen that only the F=0-1
components are clear of any complication.
%Each species shows a primary absorption component at +64~km~s$^{-1}$.  Also,
%\MtCN, \EtCN, and \CYCN\ show emission at +73~km~s$^{-1}$, and \VyCN, \EtCN\ and \CYCN\ show
%absorption at +82~km~s$^{-1}$.  
%This velocity structure is consistent with the large molecules propenal (CH$_2$CHCHO), propanal (CH$_3$CH$_2$CHO),
%and glycolaldehyde (CH$_2$OHCHO) also detected with the GBT (Hollis et al.\ 2004a). 
%The velocity structure is clearly seen in 
%the spectra of \EtCN\ and \CYCN\ but is conflicted by the hyperfine (HF) splitting present 
%in \MtCN\ and \VyCN.  For \MtCN, the absorption at +64~km~s$^{-1}$ and emission at +73~km~s$^{-1}$
%can be seen for each HF component and only absorption at +64~km~s$^{-1}$ is seen for the
%HF components of \VyCN.  The spectral line fiducial on the HF components denote the 
%primary absorption feature at +64~km~s$^{-1}$.
The panels in the right column of Figure 1 show the passbands containing
the transitions of \MtNC, \VyNC, \EtNC\ and \CYNC.  We clearly detect
the absorption and emission components at +64, +73 and +82~km~s$^{-1}$ from \MtNC\ and 
there is possible narrow emission from \VyNC\ near +64~km~s$^{-1}$.  The primary 
absorption component of \MtNC\ at +64~km~s$^{-1}$ is more than an 
order of magnitude weaker in intensity than the corresponding feature of \MtCN, however
this is the first interstellar report of this line. No emission or 
absorption is seen from either \EtNC\ or \CYNC\ beyond a 1~$\sigma$ detection limit ($\sim$~2~mK).

The $J_K=4_K-3_K$ ($K=0-3$) \MtNC\ observations using the Berkeley-Illinois-Maryland 
Association (BIMA)
Millimeter Array\footnote{Operated by the University of California, Berkeley, the 
University of Illinois, and the University of Maryland with support from the National Science
Foundation.} were made in 2001 April 20 in its C configuration toward Sgr~B2(N-LMH).  The pointing
position was the same as used for the GBT. Table 2 lists the 
molecular line parameters of each species in the \MtNC\ passband.
The species, the calculated rest frequencies, the transition quantum 
numbers, the product of the line strength and the square of the dipole
moment ($S\mu^2$), and the upper level energies are listed in the first five columns of Table 2.  
The synthesized beamwidth for these observations is $\theta_{a}$$\times$$\theta_{b}$~=~23.$''$9$\times$7.$''$1 
corresponding to a Gaussian beamsize of $\Omega_B$=192~arcsec$^2$.  
The quasar~1733-130 was used to calibrate the antenna based gains and Uranus was used 
as the flux density 
calibrator.  The absolute amplitude calibration of these sources is accurate to 
within $\sim$20\%.  The passbands were calibrated automatically during data
acquisition; a technical description of auto-calibration can be found at
http://astron.berkeley.edu/$\sim$plambeck/technical.html.
The spectral window containing the \MtNC\ transitions
has a bandwidth of 50 MHz and is divided into 128 channels for a
spectral resolution of 0.39 MHz per channel.  All data were combined,
imaged, and self-calibrated using the MIRIAD software package (Sault,
Teuben \& Wright 1995). 

Figure 2 shows fiducials for the $J_K=4_K-3_K$ ($K=0-3$) transitions of \MtNC\
in a passband dominated by strong emission from \EtCN\ which is largely confined
to the LMH hot core (Liu \& Snyder 1999).
The rest frequency located in the upper left of the panel corresponds to the 
$J_{K_a,K_c}=9_{28} - 8_{27}$ transition of \EtCN\ for an assumed LSR velocity of +64 km~s$^{-1}$.
Dashed lines denote LSR velocities of +64, +73 and +82 km~s$^{-1}$.  
The transition of \EtCN\ clearly shows strong emission components at the LMH systemic velocity of
+64 km~s$^{-1}$ and also at +73 km~s$^{-1}$.
The $J_K = 4_K - 3_K$ ($K=0 - 3$) \MtNC\ transitions are labeled in this passband
assuming an LSR velocity of +64 km~s$^{-1}$ and, except for the $J_K = 4_3 - 3_3$ transition, 
all lie in a region that is clear from
any other molecular line emission.  However, no transition of \MtNC\ is detected above a
1 $\sigma$ detection limit ($\sim$0.3 Jy beam$^{-1}$).  The 1 $\sigma$ rms noise level fiducial 
is shown at the left of the panel.

\section{DISCUSSION}

We have presented two disparate sets of observations:  GBT single
antenna results with a Gaussian beamsize of $\Omega_B$=1723~arcsec$^2$, and BIMA array
results obtained with a Gaussian beamsize of $\Omega_B$=192~arcsec$^2$.  Given the GBT beamsize 
is a factor of 9 greater than the BIMA beamsize, the
spatial scale of the GBT will couple better to the extended emission toward
Sgr~B2(N).  Conversely, the BIMA array will tend to couple better
to the 28 arcsec$^2$ ($\sim$5$''$ diameter) of the LMH hot core.  Hence, we expect that the GBT
observations will be characterized by a lower excitation temperature (T$_{ex}$) than the
excitation temperature associated with the BIMA observations.  First, we will 
derive the cyanide and isocyanide column densities from the GBT observations.
Then, we will show from BIMA observations that \MtNC\ must be extended.

%using the column 
%density of \MtCN\ determined  we derive the 
%expected integrated line intensity of \MtNC\ from the BIMA observations using an 
%excitation temperature of 170 K (Pei, Liu \& Snyder 2000) and assuming the
%\MtCN\ and \MtNC\ emission are co-spatial.  

For the GBT observations, following the formalism outlined in Hollis et al.\ (2004b),
the total beam-averaged column density of a molecular species obtained with a single element
radio telescope is given as:

For emission:
 
\begin{equation}
N_T = (1.8\times10^{14}) \frac{Q_r  \hspace{0.1cm}e^{E_u/T_{ex}}}{\nu S \mu^2} \frac{\left\{ \frac{\Delta T_A^* \Delta V}{\eta_B} \right\}}{\left\{ 1-\frac{(e^{(4.8\times10^{-5})\nu/T_{ex}}-1)}{(e^{(4.8\times10^{-5})\nu/T_{bg}}-1)} \right\}}  \hspace{0.3cm} {\rm cm^{-2}}.
\end{equation}

For absorption:

\begin{equation}
N_T = (8.5\times10^9) \frac{Q_r \left\{ \frac{\Delta T_A^* \Delta V}{\eta_B} \right\}}{\left\{ T_{ex} - \frac{T_c}{\eta_B} \right\} S \mu^2 \left\{ e^{-E_l/T_{ex}}-e^{-E_u/T_{ex}} \right\}}  \hspace{0.3cm} {\rm cm^{-2}}.
\end{equation}

In both equations, the line shapes are assumed to be Gaussian; $\eta_B$ is the telescope beam
efficiency; T$_{ex}$ is the excitation temperature; $\Delta T_A^*\Delta V$ is the product of the fitted  
line intensity (mK) and line width (km~s$^{-1}$); $Q_r$ is the rotational partition function; 
$S\mu^2$ is the product of the transition line strength and the square of the dipole 
moment (Debye$^2$); and $E_u$ is the upper rotational energy level (K).  In equation (1), 
$\nu$ is the transition frequency (MHz) and $T_{bg}\sim$2.7 K is the cosmic background 
temperature.  In equation (2), $T_c$ is the   source continuum temperature (K), and 
$E_l$ is the lower rotational energy level (K).

Using the GBT at K-band, Hollis et al.\ (2004b) found a glycolaldehyde excitation 
temperature of $T_{ex}$=8~K that satisfied the detected integrated line intensities of the absorption 
components at +64 and +82~km~s$^{-1}$ and the emission 
component at +73~km~s$^{-1}$.  Since the LSR velocity components of the cyanide and 
isocyanide molecules in Figure 1 appear to be similar, we used $T_{ex}$=8~K 
to find the total column density of each of the detected species.  Columns (7) to (12) of Table 1 summarize
the Gaussian fitting intensities and line widths for each detected transition
or 3 $\sigma$ upper limits for the intensities of non-detections at LSR velocities of +64, +73 and +82~km~s$^{-1}$.

In order to determine the intensity and line width of the individual HF and velocity components of the 
CH$_3$CN and CH$_2$CHCN lines, multiple Gaussian fits were used.  By fixing the frequency separation
between HF components, we fit the +64, +73 and +82~km~s$^{-1}$ velocity components for each HF 
component.  If the fitting routine found a solution for the intensity of a velocity component less than the 3 $\sigma$ rms
noise level, it was omitted, and the fit recalculated for one less velocity component.  The results are listed in columns 7-12 of
Table 1.  In Figure 1a, the solid line is the result of Gaussian fitting the 3 HF structure components
to the +64, +73 and +82~km~s$^{-1}$ LSR velocity structure (see Table 1).
Furthermore, as a residual to this methodology, the small absorption dip near 50 km~s$^{-1}$ in Figure 1a is not due
to a HF component nor is it a velocity component of CH$_3$CN.

The column densities listed in column (13) of Table 1 are the sum of all detected velocity components, or, 
in the cases of \MtCN\ and \VyCN, of all velocity and HF components.  The column densities of the cyanide 
molecules are within a factor of 3 or 4 of the \MtCN\ column density of ($\sim$10$^{14}$ cm$^{-2}$).
Furthermore, the column density measured from the +64~km~s$^{-1}$ velocity component accounts for $\sim$75\% 
of the total CH$_3$CN column density. The +64~km~s$^{-1}$ velocity component 
is often the only one resolved at higher frequencies and is the basis for the majority of CH$_3$CN column
density measurements at 1 and 3 mm wavelengths.

The similarity in column densities between cyanides is also seen at higher resolution toward other sources 
(Remijan et al.\ 2004b).  The  column density ratio CH$_3$NC/CH$_3$CN is $\sim$0.02
and is relevant if and only if the two species are co-spatial.  The
possible detection of \VyNC\ gives a column density ratio CH$_2$CHNC/CH$_2$CHCN of
$\sim$0.005.  
%Our CH$_3$NC/CH$_3$CN column density ratio is consistent with 
%previously determined upper limits (DeFrees, McLean \& Herbst 1985; Irvine \& Schloerb 1984).

From Remijan et al.\ (2003), the total beam averaged column density   
obtained with an interferometer is:

\begin{equation}
N_{T} = 2.04\times \frac{\int \Delta I dv}{B \theta_{a}\theta_{b}} \frac{Q_{r} \hspace{0.1cm} e^{(E_{u}/T_{ex})}} {\nu^{3} S \mu^{2}}\times10^{20} \hspace{0.3cm} {\rm cm^{-2}}.
\end{equation}

\noindent In equation (3), $\int\Delta I dv$ is the product of the
fitted line intensity (Jy beam$^{-1}$) and line width (km~s$^{-1}$), 
$\theta_{a}$ and $\theta_{b}$ are the FWHM beam widths (arcsec) and $B=\Omega_S/[\Omega_B + \Omega_S]$ is the
beam filling factor where $\Omega_B$ is the solid angle subtended by the synthesized 
beam of the interferometer and $\Omega_S$ is the solid angle subtended by the source emission. 
For the BIMA array data, $\theta_{a}$$\times$$\theta_{b}$ = 23.$''$9$\times$7.$''$1.
Columns (7) to (12) of Table 2 summarize
the Gaussian fitting intensities and line widths for \EtCN\
and the 3 $\sigma$ upper limits for the intensities of the non-detections
at LSR velocities of +64, +73 and +82~km~s$^{-1}$. The column density of \EtCN\ in 
column (13) is based on the excitation temperature of methanol (CH$_3$OH) of 170 K (Pei, Liu \& Snyder 2000).

No emission for the $J = 4-3$ $(K=0-3)$ transition of \MtNC\ was seen in the
BIMA array SgrB2(N) data.  However, if \MtNC\ and \MtCN\ are co-spatial
(i.e., our GBT CH$_3$NC/CH$_3$CN ratio of $\sim$0.02 is valid), then observations
of CH$_3$CN by other investigators which yield large CH$_3$CN column densities
predict that we should have detected CH$_3$NC with the BIMA Array if
the source is compact.  In an extensive survey encompassing the $J=12\rightarrow11$, 
$J=13\rightarrow12$ and $J=14\rightarrow13$ transitions of CH$_3$CN at 1 mm wavelengths, 
Nummelin et al.\ (2000) obtained a CH$_3$CN beam averaged column density of 
$\sim$4.2$\times$10$^{15}$~cm$^{-2}$ for a 352 K LMH hot core temperature. 
These observations were performed using the 15m SEST telescope with a $\sim$600 arcsec$^2$ 
beamsize.  Owing to the lower spatial resolution ($\sim$2 km~s$^{-1}$~channel$^{-1}$) at 1 mm
wavelengths and the large line widths (15-20 km~s$^{-1}$) of CH$_3$CN lines toward
Sgr B2(N-LMH) (Nummelin et al.\ 2000), it is not possible to resolve 
the individual velocity components of the CH$_3$CN lines.  Thus, the
calculated column density is based solely on the +64~km~s$^{-1}$ LSR velocity component.
In addition to the the column density and temperature, Nummelin et al.\ also
found a beam filling factor of the CH$_3$CN emission toward the Sgr~B2(N) hot core
of 0.014. In their data analysis,
the column density, rotational temperature and beam filling factor were given as free parameters.
From these values, an integrated line intensity was calculated for each transition and
was compared to the observed integrated line intensity until a ``best-fit'' model was 
obtained.  The best-fit model was found by finding the minimum of the reduced $\chi^2$ 
function (Nummelin et al.\ 2000).  Assuming the model explains the physics accurately, the 
measured values of the rotational temperature, column density and beam filling factor by Nummelin at al.\
toward the Sgr~B2(N) hot core are reported at a 68.3\% confidence
interval (1 $\sigma$).  Given the $\sim$600 arcsec$^2$ beamsize of the 15m SEST telescope, 
and a beam filling factor of 0.014, the measured source size of the CH$_3$CN emitting
region toward the Sgr~B2(N) hot core is $\sim$8.5 arcsec$^2$.  This size is very similar 
to the CH$_3$CN cores toward W51e1 (7.9(6)arcsec$^2$) and W51e2 (6.0(9)arcsec$^2$) 
(Remijan et al.\ 2004a).  Furthermore, the size is also similar to the emitting regions
of \VyCN\ and \EtCN\ towards the Sgr~B2(N) hot core measured by Liu \& Snyder (1999) at
subarcsecond resolution.  Thus, $\sim$8.5 arcsec$^2$ is a reasonable measure 
of the CH$_3$CN emitting region toward the Sgr~B2(N) hot core.  Using the measured rotational 
temperature of 352 K, the scaled Nummelin et al.\ (2000) \MtCN\ column density within the 192 
arcsec$^2$ beamsize of the BIMA array is $\sim$9.9$\times$10$^{16}$~cm$^{-2}$.  Using the column 
density ratio of 0.02, the expected \MtNC\ column density of the +64~km~s$^{-1}$ LSR velocity component
should be $\sim$2$\times$10$^{15}$ cm$^{-2}$ for co-spatial isomers.

Solving equation (3) for $\int\Delta I dv$ assuming a column density of 
$\sim$2$\times$10$^{15}$ cm$^{-2}$,  we should have detected a CH$_3$NC integrated line 
intensity of $\sim$57 Jy beam$^{-1}$ km~s$^{-1}$ for the $J_K=4_0-3_0$ transition.
Assuming a typical line width of 15 km~s$^{-1}$ for the 
LMH hot core of Sgr~B2(N), the expected peak intensity of the $4_0-3_0$ transition 
is $\sim$3.8 Jy beam$^{-1}$.  However, the BIMA array data shows no CH$_3$NC emission greater than   
the 1 $\sigma$ rms upper limit of $\sim$0.3 Jy beam$^{-1}$.  The most likely explanation for the 
lack of a CH$_3$NC detection is that the emission is extended and the array cannot couple to it.  In the 
case of extended emission, the ratio of the measured line intensity, or in our case the 1 
$\sigma$ rms upper limit of $\sim$0.3 Jy beam$^{-1}$, to the expected line intensity
of $\sim$3.8 Jy beam$^{-1}$ is equal to the ratio of the solid angle subtended by the synthesized 
beam of the interferometer ($\Omega_B$) to the solid angle subtended by the source emission 
($\Omega_S$).  This is because the synthesized beam of the interferometer is only collecting 
a fraction of the flux from the extended emission region which we assume to be uniformly 
distributed.  Therefore our observations indicate that $\Omega_B$/$\Omega_S$ $<$ 0.3/3.8.  
Solving for the source size assuming a Gaussian beamsize of 192 arcsec$^2$, we find 
$\Omega_S$$>$2432 arcsec$^2$ or equivalently, a linear size of $>$100$''$(1.7$'$), which is much
larger than the typical LMH hot core diameter of $\sim$5$''$ (Hollis et al.\ 2003). 
This strongly suggests that the spatial distribution of \MtNC\ is extended and would not 
show any significant concentration toward the LMH hot core.

DeFrees, McLean, \& Herbst (1985) theoretically calculated a CH$_3$NC/CH$_3$CN
abundance ratio for dense interstellar clouds in the range of 0.1-0.4 by
assuming that both species are formed from ion precursors and destroyed by
gas phase, ion molecule reactions.  The DeFrees et al.\ calculations show
that when CH$_3$$^+$ collides with HCN, a collision complex is formed that
rapidly equilibrates to protonated methyl isocyanide (CH$_3$NCH$^+$) and
protonated methyl cyanide (CH$_3$CNH$^+$) due to the large amount of energy produced from the initial
collision.  As the ensemble of ions continues to
relax, the internal energy of CH$_3$CNH$^+$ becomes insufficient to overcome the
barrier to isomerize to CH$_3$NCH$^+$.  The end result finds CH$_3$CN formed
preferrentially over CH$_3$NC.  We performed ab initio calculations in order to quantify the energy differences between
cyanide and isocyanide isomers in this work. 
Table 3 lists each molecular species (col.\ 1), 
the zero-point corrected bonding energy (ZPCBE in col.\ 2) and the relative ZPCBE (col.\ 3).
Column (2) ZPCBE represents the complete dissociation of the molecule into its constituent atoms.
%All energies are given in units of cm$^{-1}$
%and the dipole measurements are given in Debye.  
Note that the relative bonding energies between isomers are very large ($>$8500 cm$^{-1}$).  
Thus, the cyanide molecules are much more stable than their
isocyanide isomers, and therefore it is highly unlikely that any thermal process can satisfactorily 
explain the formation of isocyanide molecules.
This indicates that a non-thermal process is necessary to convert a cyanide into an isocyanide species.

In recent laboratory experiments performed by Hudson \& Moore (2004), cyanide species were 
bombarded by both protons and UV-photons to determine what would happen to the structure of the molecule.
Bombarding a pure ice sample of \MtCN\ at 24 K, the proton experiment yielded ketenimine (CH$_2$CNH),
another isomer of \MtCN, as well as \MtNC\ in a slightly lower abundance.
Similarly, the UV-photon experiment at 12 K yielded \MtNC\ 
as the primary product.  These products were also seen when \MtCN\ was embedded in N$_2$ ice
and then irradiated.  Hudson \& Moore (2004) also note that the temperature in both experiments
did not affect the formation of the end products.  In an UV-photon experiment performed at 100 K, the
destruction of \MtCN\ was faster than at 12 K, but the same end products were seen.  
This suggests that the reactions were driven more by the radiation dose 
than the sample temperature (Hudson \& Moore 2004), reinforcing 
our suggestion that non-thermal processes may be the primary route to the formation of interstellar isocyanides.  
Laboratory experiments involving \VyCN, \EtCN, and cyanoacetylene (HC$_2$CN), all showed similar results for
isocyanide isomers.  It is likely that a similar reaction involving \CYCN\ will produce \CYNC.  Furthermore, these laboratory experiments
also suggest that all isocyanides are likely to be formed under
the conditions which interstellar \MtNC\ was detected.  Based on the column density ratios and
upper limits in Table 1, we did not detect any emission from
\EtNC\ and \CYNC\ because of the possibility that we did not integrate long enough
to detect these species.
%Indeed, if the typical column density ratio of a
%cyanide to isocyanide is $\sim$0.10 as measured by our observations of \MtCN\ and \MtNC {\it and}
%the cyanide and isocyanide are co-spatial, then the emission from \VyNC\ should be $\sim$8 mK
%with respect to the same transition of \VyCN, and the emission of \EtNC\ should be $\sim$3 mK
%for the 2$_{11}$-1$_{10}$ transition of \EtNC.  Unfortunately we observed the 2$_{12}$-1$_{11}$
%transition of \EtNC\ which has a larger $K_c$ value and thus, may be less populated.
%Nonetheless, we did not integrate on source long enough
%to detect either of these species at more than the 1 $\sigma$ level ($\sim$5 mK).  The same
%assumption for \CYNC\ should have yielded emission at the 10 mK level.  This is the only 
%set of data where we integrated for 2 full tracks, reducing the noise level down to 
% $\sim$3 mK rms.  The fact that we do not see any emission from \CYNC\ may imply that: (1) 
%\CYCN\ may have a different distribution than the other cyanides in this work; (2) \CYCN\  
%may be depleted before forming its isocyanide; or (3) because the ZPCBED of \CYCN\ is almost 2000 cm$^{-1}$
%above the ZPCBED of \MtNC, it may not form in the same abundance ratio compared to the
%other cyanides.  As a possible 
%explanation, 
On the other hand, Hudson \& Moore (2004) also conducted
experiments with cyanides embedded in H$_2$O ice and found that hydrogen addition and subtraction 
reactions were taking place in both radiation bombardment reactions in which {\it no} isocyanides 
were formed.  In these experiments HC$_2$CN hydrogenated to \VyCN\ and then
to \EtCN. Based on the aforementioned laboratory results, we will suggest how large 
cyanides and their corresponding isocyanide isomers may be produced in the extended regions 
surrounding Sgr~B2(N).

The Sgr~B2(N) region is known to contain widespread shocks (Chengular \& Kanekar 2003).  
Such shocks have been presumed to be responsible for the formation
and distribution of large aldehyde molecules toward this region (Hollis et al. 2004a, b).  
Furthermore, it is also well known that there is ambient UV flux and cosmic ray flux in 
these regions.  On the surfaces and mantles of dust grains, larger species such as CH$_3$CN and HC$_2$CN
may be present. \VyCN\ and \EtCN\ are formed as a result of H-addition reactions, thus
depleting HC$_2$CN.  As a shock passes, these large molecules are expelled from the 
grain surface into the gas phase. At this point UV and cosmic rays may interact with the cyanide species and 
form isocyanides.   As a result, the isocyanides show an extended distribution.  
In hot cores, there is a lower UV and cosmic ray flux due to the large column density of gas and dust. 
Furthermore, thermal processes are the primary mechanism to liberate molecules off grain
surfaces (Tielens \& Hagen 1982; Hasegawa, Herbst, \& Lueng 1992).  Indeed, this is initially why 
these regions were believed to be the ideal places to form large molecules (Ehrenfreund \& Charnley 2000).  
Also, in these compact regions, it is believed warm gas phase chemistry in addition to grain
surface chemistry can form large molecules which may account for the enhanced abundances 
of \VyCN\ and \EtCN\ in hot core regions (Liu \& Snyder 1999, see also \EtCN\ column density in Table~2 
with respect to Table~1). However, thermal processes are not enough to form isocyanide
species after the cyanides have been expelled from the grain and hence there should be a 
depletion of isocyanides towards hot cores.  Our observations support this hypothesis.

In summary, we have calculated the chemical bonding energies for cyanide
molecules studied herein (\MtCN, \VyCN, \EtCN, \& \CYCN) and found they are 
much greater ($>$8500 cm$^{-1}$) than that of the corresponding isocyanide isomers.  As a consequence, cyanide
interstellar molecules are most likely preferentially formed at the
expense of isocyanide isomers.  Moreover, in the case of CH$_3$CN, it is seen in
the LMH hot core and in the surrounding medium while CH$_3$NC appears to be
deficient in the LMH hot core.  This suggests that large-scale,
non-thermal processes such as shocks or enhanced UV flux in the surrounding medium account
for the conversion of CH$_3$CN to CH$_3$NC while the LMH hot core is dominated
by thermal processes which cannot produce a significant amount of CH$_3$NC.  Hudson \& Moore
(2004) have shown that radiation bombardment of CH$_3$CN in ice analog
experiments can produce CH$_3$NC, thus tending to support what is observed.
Additionally, since CH$_3$NC is deficient in the LMH and CH$_3$CN is
prevalent, this suggests that the relative abundance ratio between these
two isomers is not meaningful because the molecules are not co-spatial. 
This may well apply to other cyanide and isocyanide isomers.

J.M.H.\ gratefully acknowledges research support from H.A.\ Thronson, Assistant Associate Administrator
for Technology, NASA Science Mission Directorate. We also greatly thank an anonymous referee whose comments 
and suggestions provided additional insight to this manuscript.

\clearpage

\begin{deluxetable}{lccccccccccccc}
\tabletypesize{\tiny}
\tablewidth{43.5pc}
\tablecolumns{14}
\tablecaption{GBT Observations - Cyanide and Isocyanide Molecular Line Parameters and Column Densities}
\tablehead{
\colhead{Species} & \colhead{Frequency} & \colhead{Transition} & \colhead{S} & \colhead{E$_u$} & \colhead{T$_c$} & \multicolumn{2}{c}{(+64 km~s$^{-1}$)} & \multicolumn{2}{c}{(+73 km~s$^{-1}$)} & \multicolumn{2}{c}{(+82 km~s$^{-1}$)} & \colhead{N$_{\rm T}$}& \colhead{\underline{N$_{\rm T}$(XNC)}}\\
\cline{7-12}
\colhead{} & \colhead{} & \colhead{} & \colhead{} & \colhead{} & \colhead{} & \colhead{($\Delta T_A^*$)} & \colhead{($\Delta V$)} & \colhead{($\Delta T_A^*$)} & \colhead{($\Delta V$)} & \colhead{($\Delta T_A^*$)} & \colhead{($\Delta V$)} & \colhead{$\times10^{-13}$} & \colhead{N$_{\rm T}$(XCN)}\\
\colhead{} & \colhead{(MHz)} & \colhead{} & \colhead{} & \colhead{(K)} & \colhead{(K)} & \colhead{(mK)} & \colhead{(km/s)} & \colhead{(mK)} & \colhead{(km/s)} & \colhead{(mK)} & \colhead{(km/s)} & \colhead{(cm$^{-2}$)} & \colhead{}\\
\colhead{(1)} & \colhead{(2)} & \colhead{(3)} & \colhead{(4)} & \colhead{(5)} & \colhead{(6)} & \colhead{(7)} & \colhead{(8)} & \colhead{(9)} & \colhead{(10)} & \colhead{(11)} & \colhead{(12)} & \colhead{(13)} & \colhead{(14)}\\
}
\startdata
\MtCN\tablenotemark{a} & 18396.7252(7) & 1$_0$-0$_0$ & 1.332 & 0.88 & 10.6 & -275(25) & 5.8(7) & 160(20) & 7.9(9) &  $<$7 & ... & & \\
 &  &  $F=1-1$ & & & & & & & & & & &\\
 & 18397.9965(6) & 1$_0$-0$_0$  & 2.220 & 0.88 & 10.6 & -400(15) & 7.0(3) & 310(19) & 5.9(4) &  $<$7 & ... & 10.8(20) & \\
 &  &  $F=2-1$ & & & & & & & & & & & \\
 & 18399.8924(3) & 1$_0$-0$_0$  & 0.444 & 0.88 & 10.6 & -180(25) & 8.3(9) & $<$7 & ... & $<$7 & ... & & 0.02(2) \\
 &  &  $F=0-1$ & & & & & & & & & & & \\
\MtNC\tablenotemark{b} & 20105.754(1) & 1$_0$-0$_0$ & 4.000 & 0.88 & 10.2 & -48(2) & 8.4(4) & 11(3) & 6.4(9) & $<$7 & ... & 0.2(1) & \\
 &  &  & & & & & & & & &  & &\\
\hline
 &  &  & & & & & & & & &  & &\\
\VyCN\tablenotemark{c} & 18512.158(13) & 2$_{12}$-1$_{11}$  & 0.375 & 3.51 & 11.1 & -62(2) & 4.9(1) & $<$7 & ... & $<$7 & ... &  &\\
 &  &  $F=2-1$ & & & & & & & & &  & &\\
 & 18513.311(7) & 2$_{12}$-1$_{11}$  & 0.701 & 3.51 & 11.1 & -82(2) & 4.9(1) & $<$7 & ... & $<$7 & ... & 37.2(30) &\\
 &  &  $F=3-2$ & & & &  & & & & & & &\\
 & 18514.400(27) & 2$_{12}$-1$_{11}$  & 0.167 & 3.51 & 11.1 & -19(3) & 4.1(5) & $<$7 & ... & $<$7 & ... & & 0.005(2)\\
 &  &  $F=1-0$ & & & &  & & & & & & &\\
\VyNC\tablenotemark{d} & 19993.916(1) & 2$_{12}$-1$_{11}$ & 1.500 & 3.67 & 10.5 & 8(2) & 1.5(4) & $<$7 & ... & $<$7 & ... & 0.2(5) &\\
 &  &  & & & & & & & & & & &\\
\hline
 &  &  & & & & & & & & & & & \\
\EtCN\tablenotemark{e} & 18377.721(3) & 2$_{11}$-1$_{10}$ & 1.500 & 1.56 & 10.0 & -24(2) & 2.8(2) & 20(2) & 2.8(2) & -20(2) & 2.8(2) & 3.5(4) &\\
 &  &  & & & & & & & & & & & $<$0.3\\
\EtNC\tablenotemark{f} & 18803.209(10) & 2$_{12}$-1$_{11}$ & 1.500 & 1.56 & 9.8 & $<$7 & ... & $<$7 & ... & $<$7 & ... & $<$1.1 &\\
 &  &  & & & & & & & & & & &\\
\hline
 &  &  & & & & & & & & & & &  \\
\CYCN\tablenotemark{g} & 18638.617(2) & 7-6 & 7.000 & 2.70 & 10.9 & -63(2) & 5.8(2) & 97(1) & 16.0(5) & -59(2) & 8.0(3) & 6.1(2) &\\
 &  &  & & & &  & & & & & & & $<$0.03\\
\CYNC\tablenotemark{h}  & 19616.504(2) & 7-6 & 7.000 & 2.84 & 9.7 & $<$5 & ... & $<$5 & ... & $<$5 & ... & $<$0.2 &\\

\enddata
\tablecomments{Uncertainties in the frequencies refer to the least significant digit (Taylor \& Kuyatt 1994) and 
are 2 $\sigma$ values (coverage factor 2). Upper limits of the line intensities are 3 $\sigma$ values.  All 
other listed uncertainties are 1 $\sigma$.}
\tablenotetext{a}{Rest Frequencies from Boucher et al.\ 1980; $\mu_a$=3.922(1) from Gadhi et al.\ 1984}
\tablenotetext{b}{Rest Frequencies from Kukolich 1972;  $\mu_a$=3.830(60) from Ghosh et al.\ 1953}
\tablenotetext{c}{Rest Frequencies from Gerry et al.\ 1979; $\mu_a$=3.815(12), $\mu_b$=0.894(68) from Stolze, M. \& Sutter D. H. 1985}
\tablenotetext{d}{Rest Frequencies from Yamada \& Winnewisser 1975; $\mu_a$=3.470(30), $\mu_b$=0.790(100) from Bolton, Owen, \& Sheridan 1970}
\tablenotetext{e}{Rest Frequencies, $\mu_a$=3.850(10) and $\mu_b$=1.230(60) from Lovas 1982}
\tablenotetext{f}{Rest Frequencies from Fliege \& Driezler 1985; $\mu_a$=3.790(20) and $\mu_b$=1.310(40) from Anderson \& Gwinn 1976}
\tablenotetext{g}{Rest Frequencies from Winnewisser et al.\ 1982; $\mu_a$=4.330(30) from Alexander et al.\ 1976}
\tablenotetext{h}{Rest Frequencies from Botschwina et al.\ 1998; $\mu_a$=3.590 calculated in this work}

\end{deluxetable}

\begin{deluxetable}{lccccccccccc}
\tabletypesize{\tiny}
\tablewidth{40pc}
\tablecolumns{12}
\tablecaption{BIMA Array Observations - Molecular Line Parameters and Column Densities}
\tablehead{
\colhead{Species} & \colhead{Frequency} & \colhead{Transition} & \colhead{S$\mu^2$} & \colhead{E$_u$} & \multicolumn{2}{c}{(+64 km~s$^{-1}$)} & \multicolumn{2}{c}{(+73 km~s$^{-1}$)} & \multicolumn{2}{c}{(+82 km~s$^{-1}$)} & \colhead{N$_{\rm T}$}\\
\cline{6-11}
\colhead{} & \colhead{} & \colhead{} & \colhead{} & \colhead{} & \colhead{($\Delta I$)} & \colhead{($\Delta V$)} & \colhead{($\Delta I$)} & \colhead{($\Delta V$)} & \colhead{($\Delta I$)} & \colhead{($\Delta V$)} & \colhead{} \\
\colhead{} & \colhead{(MHz)} & \colhead{} & \colhead{(Debye$^2$)} & \colhead{(K)} & \colhead{(J/bm)} & \colhead{(km/s)} & \colhead{(J/bm)} & \colhead{(km/s)} & \colhead{(J/bm)} & \colhead{(km/s)} & \colhead{(cm$^{-2}$)} \\
\colhead{(1)} & \colhead{(2)} & \colhead{(3)} & \colhead{(4)} & \colhead{(5)} & \colhead{(6)} & \colhead{(7)} & \colhead{(8)} & \colhead{(9)} & \colhead{(10)} & \colhead{(11)} & \colhead{(12)}\\
}
\startdata
H$_2$CCCC\tablenotemark{a} & 80383.96(2) & 9$_{09}$-8$_{08}$ & 153.8 & 19.3 & $<$0.3 & ... & $<$0.3 & ... & $<$0.3 & ... & $<$6.0$\times$10$^{14}$\\
HCOOCH$_3$\tablenotemark{b} & 80395.135(126) & 9$_{28}$-9$_{09}$ E & 0.7 & 28.8 & $<$0.3 & ... & $<$0.3 & ... & $<$0.3 & ... & $<$4.7$\times$10$^{17}$\\
\EtCN & 80404.898(12) & 9$_{28}$-8$_{27}$ & 126.1 & 23.8  & 4.6(1) & 16.0(8) & 1.7(3) & 5.1(6) & 0.6(2) & 13.6(4.0) & 3.1$\times$10$^{16}$\\
\MtNC & 80405.504(120) & 4$_{3}$-3$_{3}$ & 29.6 & 73.0 & $<$0.3 & ... & $<$0.3 & ... & $<$0.3 & ... &\\
\MtNC & 80414.660(120) & 4$_{2}$-3$_{2}$ & 50.7 & 37.4 & $<$0.3 & ... & $<$0.3 & ... & $<$0.3 & ... & \\
\MtNC & 80420.062(120) & 4$_{1}$-3$_{1}$ & 63.3 & 16.0 & $<$0.3 & ... & $<$0.3 & ... & $<$0.3 & ... &\\
\MtNC\tablenotemark{c} & 80421.910(120) & 4$_{0}$-3$_{0}$ & 67.6 & 8.9 & $<$0.3 & ... & $<$0.3 & ... & $<$0.3 & ... & $<$4.0$\times$10$^{13}$\\
\enddata
\tablenotetext{a}{Rest Frequency and Line Strength (S$\mu^2$) from Killian et al.\ 1990}
\tablenotetext{b}{Rest Frequency and Line Strength (S$\mu^2$) from Bauder 1979}
\tablenotetext{c}{$N_T$ calculated using the strongest expected component.}
\end{deluxetable}

\begin{deluxetable}{ccc}
\tabletypesize{\scriptsize}
\tablewidth{15pc}
%\tablecolumns{3}
\tablecaption{Chemical Bonding Energy}
\tablehead{
\colhead{Species} & \colhead{ZPCBE\tablenotemark{a}} & \colhead{Relative} \\
\colhead{} & \colhead{} & \colhead{ZPCBE} \\
\colhead{} & \colhead{(cm$^{-1}$)} & \colhead{(cm$^{-1}$)} \\
\colhead{(1)} & \colhead{(2)} & \colhead{(3)} \\
}
\startdata
\MtCN & -29,070,027 & 0 \\
\MtNC & -29,060,541 & 9,486 \\
\VyCN & -37,408,030 & 0 \\
\VyNC & -37,399,372 & 8,658 \\
\EtCN & -37,672,368 & 0 \\
\EtNC & -37,663,671 & 8,697 \\
\CYCN & -53,818,766 & 0 \\
\CYNC & -53,807,547 & 11,219 \\

\enddata
\tablecomments{The Chemical Bonding Energies were Obtained at the MP2/aug-cc-pVTZ Level of Theory (Frisch et al.\ 2004; Dunning 1989;
Kendall, Dunning \& Harrison 1992; Woon \& Dunning, 1993)}
\tablenotetext{a}{Zero-Point Corrected Bonding Energy}

\end{deluxetable}

\clearpage

\figcaption{GBT cyanide and isocyanide spectra toward SgrB2(N) at 24.4 kHz channel spacing.  
Transition quantum numbers are shown in each panel.  Each spectrum was processed with a median filter 
to remove instrumental slopes in the bandpass.  The abscissa is the 
radial velocity with respect to the LSR calculated for the rest frequency of the transition shown 
(see Table 1) at an assumed source velocity of +64 km~s$^{-1}$.  Dashed lines show LSR velocities at 
+64, +73 and +82 km~s$^{-1}$.  In the case of \MtCN\ and \VyCN\, the HF structure is also seen and 
labeled for an LSR velocity of +64 km~s$^{-1}$.}

\figcaption{BIMA array spectral passband toward the LMH hot core of Sgr~B2(N) containing the $J_K=4_K-3_K$ ($K=0-3$) transitions of 
\MtNC.  In this case, the rest frequency corresponds to the \EtCN\ transition at 80.4049 GHz for
a LSR velocity of +64 km~s$^{-1}$.  However, there are also emission components at +73
and +82 km~s$^{-1}$.  Finally, the locations of the transitions of \MtNC, HCOOCH$_3$ and H$_2$CCCC
(Table 2) are also labeled assuming an LSR velocity of +64 km~s$^{-1}$.}

\clearpage

\begin{figure}
\epsscale{0.9}
\plotone{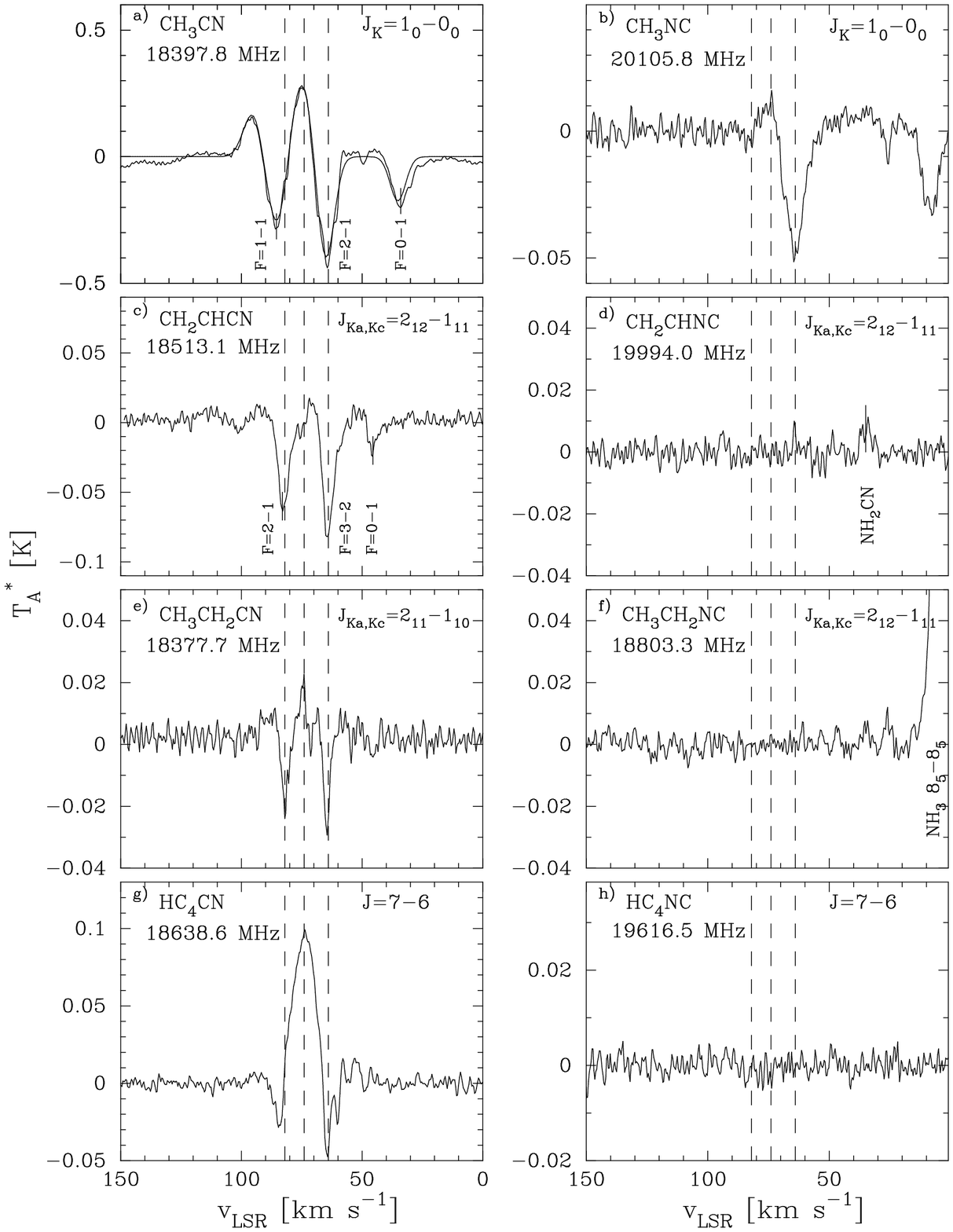}
\centerline{Figure 1.}
\end{figure}

\begin{figure}
\epsscale{1.0}
\plotone{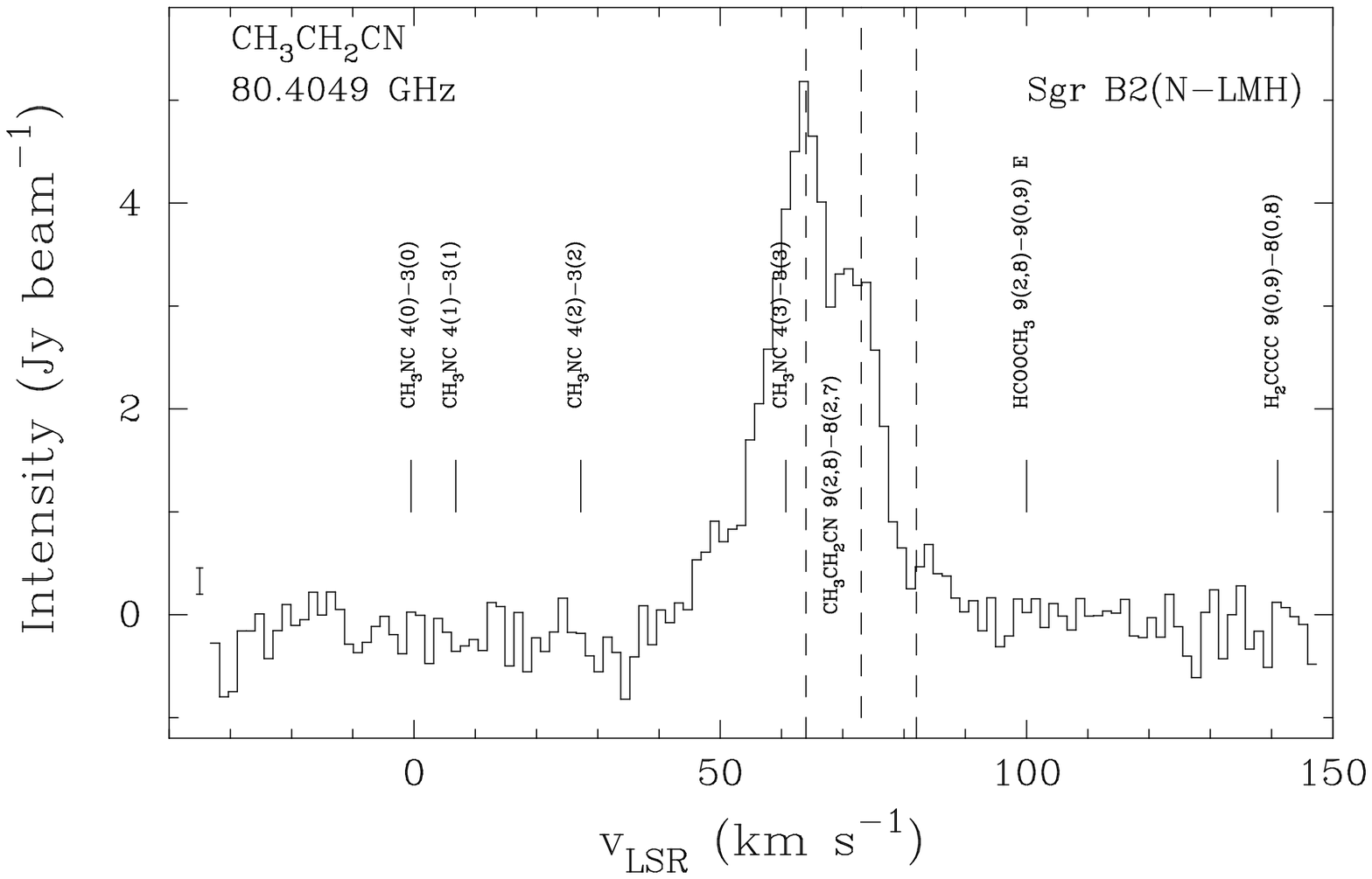}
\centerline{Figure 2.}
\end{figure}


\begin{thebibliography}{}

\bibitem[]{450}Alexander, A. J., Kroto, H. W., \& Walton, D. R. M. 1976, J. Mol. Spect., 62, 175
 
\bibitem[]{452}Anderson, R. J. \& Gwinn, W. D. 1968, J. of Chem. Physics, 49, 3988

\bibitem[]{454}Bauder, A. 1979, J. Phys. Chem. Ref. Data, 8, 583

\bibitem[]{456}Bolton, K., Owen, N. L., \& Sheridan, J. 1970, Spectrochim. Acta, 26A, 909

\bibitem[]{458}Botschwina, P., Heyl, A., Chen, W., McCarthy, M. C., Grabow, J.-U., Travers, M. J., \&  Thaddeus, P. 1998, J. Chem. Phys. 109, 3108

\bibitem[]{460}Boucher, D., Burie, J., Bauer, A., Dubrulle, A., \& Demaison, J. 1980, J. Phys. Chem. Ref. Data, 9, 659

\bibitem[]{462}Cernicharo, J., Kahane, C., Guelin, M., \& Gomez-Gonzalez, J. 1988, A\&A, 189, L1

\bibitem[]{464}Chengalur, J.N., \& Kanekar, N.  2003, A\&A, 403, L43

\bibitem[]{466}DeFrees, D. J., McLean, A. D., \& Herbst, E. 1985, ApJ, 293, 236

\bibitem[]{468}Dunning, T. H. Jr. 1989, J. Chem. Phys. 90, 1007

\bibitem[]{470}Ehrenfreund, P.~\& Charnley, S.~B.\ 2000, \araa, 38, 427

\bibitem[]{472}Fliege, E. \& Dreizler, H. 1985, Z. Naturforsch., 40a, 43

\bibitem[]{474}Friedel, D. N., Snyder, L. E., Turner, B. E., \& Remijan, A. 2004, ApJ, 600, 234

\bibitem[]{476}Frisch et al. 2004, Gaussian 03, Revision C.01, Gaussian, Inc., Wallingford CT

\bibitem[]{478}Gadhi, J., Lahrouni, A., Legrand, J., \& Demaison, J. 1995, J. Chim. Phys., 92, 1984

\bibitem[]{480}Gerry, M. C. L., Yamada, K., \& Winnewisser, G. 1979, J. Phys. Chem. Ref. Data, 8, 107

\bibitem[]{482}Ghosh, S. N., Trambarlo, R. \& Gordy, W. 1953, J. Chem. Phys., 21, 308

\bibitem[]{484}Hasegawa, T. I., Herbst, E., \& Leung, C. M. 1992, ApJS, 82, 167

\bibitem[]{486}Hollis, J. M., Pedelty, J. A., Boboltz, D. A., Liu, S.-Y., Snyder, L. E., Palmer, Patrick, Lovas, F. J., \& Jewell, P. R. 2003, ApJ, 596, 235

\bibitem[]{488}Hollis, J. M., Jewell, P. R., Lovas, F. J., Remijan, A., \& M$\o$llendal, H. 2004a, ApJ, 610, L21

\bibitem[]{490}Hollis, J. M., Jewell, P. R., Lovas, F. J., \& Remijan, A. 2004b, ApJ, 613, L45

\bibitem[]{492}Hudson, R. L. \& Moore, M. H. 2004, Icarus, 172, 466

%\bibitem[]{494}Irvine, W. M. \& Schloerb, F. P. 1984, ApJ, 282, 516

\bibitem[]{496}Kendall, R. A., Dunning, T. H. Jr., \& Harrison, R. J. 1992, J. Chem. Phys. 96, 6769

\bibitem[]{498}Killian, T. C., Gottlieb, C. A., Gottlieb, E. W., Vrtilek, J. M., \& Thaddeus, P. 1990, ApJ, 365, 89

\bibitem[]{500}Kukolich, S. G. 1972, J. Chem. Physics, 57, 869

\bibitem[]{502}Liu, S.-Y. \& Snyder, L. E. 1999, ApJ, 523, 683

\bibitem[]{504}Lovas, F. J. 1982, J. Phys. Chem. Ref. Data, 11, 251

\bibitem[]{506}Nummelin, A., Bergman, P., Hjalmarson, A., Friberg, P., Irvine, W. M., Millar, T. J., Ohishi, M., \& Saito, S. 1998, ApJS, 117, 427

\bibitem[]{508}Nummelin, A., Bergman, P., Hjalmarson, A., Friberg, P., Irvine, W. M., Millar, T. J., Ohishi, M., \& Saito, S. 2000, ApJS, 128, 213

%\bibitem[]{510}Pei, C.-C., Liu, S.-Y., \& Snyder, L. E. 2000, ApJ, 530, 800

\bibitem[]{512}Remijan, A., Snyder, L. E., Friedel, D. N., Liu, S.-Y., \& Shah, R. Y. 2003, ApJ, 590, 314

\bibitem[]{514}Remijan, A., Sutton, E. C., Snyder, L. E., Friedel, D. N., Liu, S.-Y., \& Pei, C. C. 2004a, ApJ, 606, 917

\bibitem[]{516}Remijan, A., Shiao, Y.-S., Friedel, D. N., Meier, D. S., \& Snyder, L. E. 2004b, ApJ, 617, 324

\bibitem[]{518}Sault, R. J., Teuben, P. J., \& Wright, M. C. H. 1995, in ASP Conf.
Ser. 77, Astronomical Data Analysis Software and Systems IV, ed. R. A. Shaw, H.
E. Payne, \& J. J. E. Hayes (San Francisco: ASP), 433

\bibitem[]{522}Stolze, M. \& Sutter, D. H. 1985, Z. Naturforsch., 40a, 998

\bibitem[]{524}Tielens, A. G. G. M., \&  Hagen, W. 1982, A\&A, 114, 245

\bibitem[]{526}Ulich, B.L., \& Haas, R.W.  1976, ApJS, 30, 247

\bibitem[]{528}Winnewisser, G., Winnewisser, M., \& Christiansen, J. J. 1982, A\&A, 109, 141 

\bibitem[]{530}Woon, D. E., \& Dunning, T. H. Jr. 1993, J. Chem. Physics, 98, 1358

\bibitem[]{532}Yamada, K. \& Winnewisser, M. 1975, Z. Naturforsch., 30a, 672

\end{thebibliography}
\end{document}